# A Novel Micro-service Based Platform for Composition, Deployment and Execution of BDA Applications

Davide Profeta, Nicola Masi, Domenico Messina, Davide Dalle Carbonare, Susanna Bonura, Vito Morreale
Engineering Ingegneria Informatica SpA
Palermo, Italy
davide.profeta@eng.it, nicola.masi@eng.it, domenico.messina@eng.it,
davide.dallecarbonare.eng.it, susanna.bonura@eng.it, vito.morreale@eng.it

*Abstract* — **Big Data are growing at an exponential rate and it becomes necessary the use of tools and technologies to manage, process and visualize them in order to extract value. In this paper a micro-service based platform is presented for the composition, deployment and execution of Big Data Analytics (BDA) application workflows in several domains and scenarios is presented. ALIDA is a result coming from previous research activities by ENGINEERING. It aims to achieve a unified platform that allows both BDA application developers and data analysts to interact with it. Developers will be able to register new BDA applications through the exposed API and/or through the web user interface. Data analysts will be able to use the BDA applications provided to create batch/stream workflows through a dashboard user interface to manipulate and subsequently visualize results from one or more sources. The platform also supports the auto-tuning of Big Data frameworks deployment properties to improve metrics for analytics application. ALIDA has been properly extended and integrated into a software solution for the analysis of large amounts of data from the avionic industries. A use case within this context is then presented.**

*Big Data analytics; Big Data systems; Spring Cloud Dataflow; Microservices; Data visualization; Visual analytics; Workflow management system; Optimization; BDA Application*

## I. INTRODUCTION

With the ever-growing Big Data, the management and use of BDA applications is becoming increasingly dominant and necessary also for business growth on different domains. This is mainly a result of cloud computing platforms that today allow industries to process, store and analyze a large amount of data on demand, reducing all the costs arising from the management and maintenance of an internal infrastructure [1] [2]. Therefore, BDA applications are increasingly to be used for model business processes instead of traditional methods based on statistical or empirical modelling, leading to lower computational effort. In the world of big data, however, these must be supported by technologies and infrastructures capable of handling a large amount of data. ALIDA has been designed for this purpose, supporting a full orchestration of BDA application workflows.

Currently, several IT solutions exist which support workflow design and orchestration. One of these is StreamSets[1]. It provides a "drag-and-drop UI for connecting sources to destinations with transformations", and allows to "build, preview, debug and schedule from a single interface". However, it is a product mainly intended to be sold and used by third parties. In fact, the only open source component does not make it usable for wider purposes in the Big Data field (e.g. a resources orchestrator like Kubernetes[2] is not supported). In the open-source market, Apache Airflow[3] allows to schedule and monitor programmatically generated analytics workflows. Nevertheless, this solution does not offer a graphical designer for the workflows creation, aiming to be used by very skilled users who know how to automate the required processes. Another solution is the one offered by Luigi[4]. It helps to build complex pipelines of batch jobs with a target-based approach. On the other hand, Luigi UI is minimal with no user interaction with running processes. Another type of orchestration is presented by Spring Cloud Data Flow[5] (SCDF). It allows to orchestrate micro-services provided as Spring Boot apps, using two different frameworks for batch and stream modes. Moreover, it offers a graphical designer for building batch/stream pipelines interactively. By providing a service package, even less skilled users can create workflows according to their needs. In conclusion, the best performing solutions are commercial; the open source solutions either are not able to manage BD workflows, or do not offer a dashboard to draw workflows. Furthermore, none of them presents a graphical user interface (GUI) for customizing and displaying result in graphs. ALIDA is a platform for composition, deployment and execution of workflows of BDA applications; such workflows may be both batch and stream. Moreover, it offers the Configuration Optimizer to improve BDA applications metrics and the Smart-Viz for result visualization.

SCDF has been chosen as the engine of the ALIDA platform. To reach the goal set by the platform, other open-source components and technologies has been used. In fact, the realized platform, using a model-based BDA-as-a-service (MBDAaaS) approach, aims to encompass all the main functionalities of the workflow management system and visualization tools under a single totally web-based environment where it is possible to create a community made up of both service providers and data analysts [3] [4].

---

[1] StreamSets - https://streamsets.com.
[2] Kubernetes - https://kubernetes.io.
[3] Apache Airflow - https://airflow.apache.org.
[4] Luigi - https://luigi.readthedocs.io/en/stable/index.html.
[5] Spring Cloud Data Flow – https://spring.io/projects/spring-cloud-dataflow.

Service providers will be able to upload their own micro-services through a GUI into a dedicated service catalogue enriched with other information. By adopting micro-services approach, modularity and re-usability are granted. This, together with a graphic designer for their composition, allows even less skilled users to create a pipeline of services (workflows) to process large amounts of data, both on batch and stream mode. The data can then be displayed and analyzed within a customizable chart via a web-based service, in order to mine meaningful information that are essential to take decisions. More and more BDA applications will be registered into the platform, thanks to their reuse and modularity, more and more users will be able to solve their tasks through a graphical orchestration of workflows, reducing development time. Moreover, using big data management technologies and the data visualization service, it is possible to extract information from a large amount of data, as well as apply machine learning analytics for different purposes: fraud detection, predictive maintenance and so on.

The rest of the paper is organized as follows. In Section II. an overview of the novel platform and their components provided. In section III. a user scenario example for the avionic industry through the ALIDA platform is shown. Lastly, in Section IV. final considerations are remarked.

## II. ALIDA PLATFORM ARCHITECTURE

ALIDA supports the composition, deployment and execution of BDA applications in several domains and scenarios, according to their specific requirements and needs with a model-based BDA-as-a-service (MBDAaaS) approach.

The ALIDA architecture, shown in Fig. 1, is micro-service oriented and exposes three interfaces: *(i)* Services Registration; *(ii)* Workflows Definition; *(iii)* Workflows Execution.

As for scalable data stores and Big Data frameworks, we took into account a number of cutting edge storage systems that cover the availability of both data streams and batch data sets, and address the variety of data formats and structures.

Among the reliable real-time data stores, we selected Cassandra to enable the ingestion of temporal data in real time and maintain these records with a long retention period, while REDIS achieves a pure in-memory ingestion of key-value records. As a workflow engine (deployment and execution), an instance of SCDF was configured in ALIDA platform.

In order to ensure a fault-tolerant and elastic deployment of the platform components, Kubernetes cluster manager has been taken into account, being also compliant with the native service deployment interface of SCDF built-in workflow engine. In ALIDA, Kafka was chosen as a distributed publish/subscribe message broker, having an implicit portability for real time and batch business applications.

An important feature provided within ALIDA platform is the implementation of a component for the auto-tuning of the Spark framework deployment properties related to the BDA applications. This optimizer synthesizes a machine learning model based on the performance metrics of the BDA application workflows running on the platform, and then provides an optimal deployment configuration, being a tuple of service tuning properties.

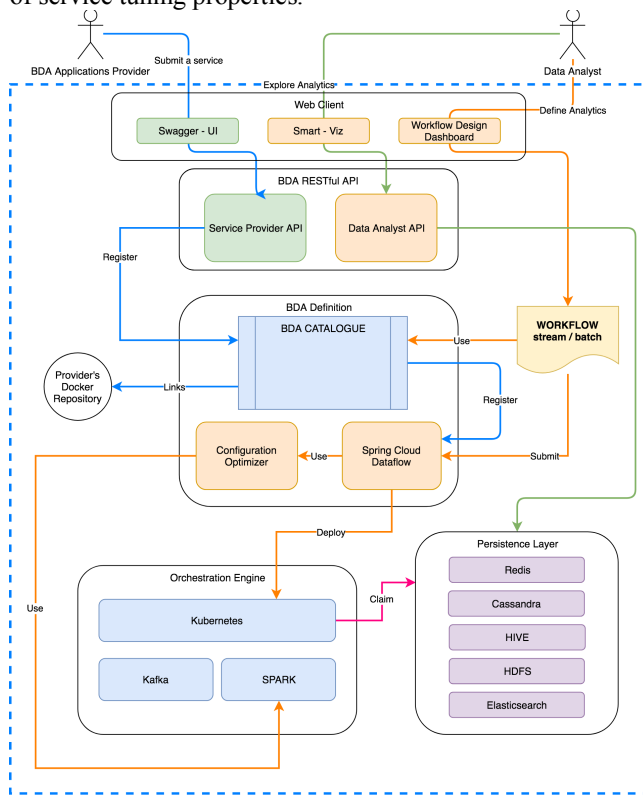

Figure 1. ALIDA general architecture.

Services registration is performed by posting a service description by the ALIDA gateway API.

Registration includes the following steps: *(i)* Store service metadata on Nexus repository as a Maven Artifact; *(ii)* Register the service on SCDF; *(iii)* Create an entry for the service to the Catalogue. In other words, the Platform Gateway API directly hits the Service Catalogue API to persist the service definition and the SCDF REST API to later compose a workflow with it. So the components involved in the services registration are the Service Catalogue, SCDF and the external repository. A GUI has been developed for the ALIDA platform, as it is can be used for exploring the services registered in the catalogue, supported frameworks, defined and running workflows. ALIDA GUI uses OAuth2[6] for registration, authentication and user profiling.

The ALIDA platform provides a web-based data visualization service called Smart-Viz. The Smart-Viz allows access, via connectors, to different data sources, displaying the databases (DBs) and tables inside them, as well as the fields of the individual tables. In addition to the default data sources it is also possible to add new ones pointing to external DBs. Most of the supported connectors are associated to NoSQL databases. Another important connector that has been added is the one related to HDFS. In fact, the most used framework to process and run analytics on a large amount of data, is Spark. It was designed to read and write data on HDFS. SQL API has been used in the HDFS connector backend in order to get data from HDFS, prepare and send them to the user on the client side.

---

[6] OAuth2 – https://oauth.net/2.

The web-based data visualization service, thanks to the provided connectors, allows an easy use for the data analysts. In fact, with no software installation on the client side, it is possible to interact with the component functionalities and result charts totally via web.

For the part of frontend it has been used Node.js, with graphic libraries D3.js for the part related to the charts. On the backend side, the different database/storage connectors are managed accordingly.

It is therefore possible to run queries and show the results on different available charts, from the simplest scatter plot to the most sophisticated dendogram. The steps following the result of the query, allow filtering and recommending of a set of charts depending on the dataset type, data cardinality, and so on. Another feature of the Smart-Viz is the visualization of streaming data from a data source through an automatic refresh of the results on the chart selected.

Finally, using this lite web-based tool, data analysts can easily extract the information and relationships from the processed data on different storages.

BDA applications running on a framework, such as Spark, are affected by the performance problem related to the configuration of the framework deployment properties. Spark contains more than 180 configuration properties that can be accurately manipulated based on application and optimization criteria. This manual configuration requires a lot of time by domain experts to optimize a specific BDA application through trials and errors.

The Configuration Optimizer of ALIDA, inspired by a work of G. Wang et al. [5] is used to optimize the execution time of a BDA-service pipeline taking full advantage of the Spark engine. All of these are benefits for the end user, especially in a BDA-as-a-service model, where there is often a pay-per-use business model based on resource and time consumption.

The Configuration Optimizer has been integrated into the ALIDA platform through four micro-services, representing the four phases on which it operates: random tunings, workload executions, performance classifier and recursive random search. These micro-services communicate according to a publish-subscribe pattern, allowing a decoupling of the components. Basically, the Configuration Optimizer uses black-box models and the recursive random search (RRS) algorithm [6] to find a sub-optimal configuration. The model's inputs are the parameter configurations, while the outputs express whether the input configuration leads to better performance on the metrics choices and how much it improves. The idea behind the Configuration Optimizer is to generate models that work well to optimize multiple services. That is, each service category is associated to a workload. A workload is an application that contains elementary operations but at the same time representative for a category of applications on which the dataset is subsequently generated to train the models, and then used to look for a sub-optimal configuration. In this way can be reused the same model for multiple services. The workload chosen for the optimization of analytics using the Spark framework executes low-level derivatives on a numerical input dataset.

### III. ALIDA IN AN AVIATION USE CASE SCENARIO

ALIDA has been properly extended and integrated in the ICARUS core platform. ICARUS is an Innovation Action (IA) project, that aims to address critical barriers for the adoption of Big Data in the aviation industry (e.g. data fragmentation, data provenance, data licensing and ownership, data veracity), and enables aviation-related Big Data scenarios for EU-based companies, organizations and scientists, through a multi-sided platform that will allow exploration, curation, integration and deep analysis of original, synthesized and derivative data characterized by different velocity, variety and volume in a trusted and fair manner. The ICARUS architecture, shown in Fig. 2, is multi-layered. The lowest layer is represented by aviation data sources. The Data Collection Services Layer is responsible for the aggregation of data from multiple sources that carry valuable aviation-related information, feeding business and public, open data into the platform, as well as for ensuring the integrity and veracity of the data. The Data Curation and Linking Services Layer will allow curation, harmonization and processing of cross-sectorial multi-lingual data to be easily consumable by various stakeholders and services. The Data Analytics Services Layer allows stakeholders to analyze and visualize results downstream of BDA applications to generate new knowledge and insights. ALIDA has been properly extended and integrated as part of the Data Analytics Service Layer, according to the ICARUS requirements and specifications.

For the use case scenario in the aviation domain, open data are taken from the *OpenFlights/Airline Route Mapper Route Database[7]*. The dataset contains routes between several airports on 548 airlines around the globe. A pipeline of BDA applications has been designed and executed in order to cluster the dataset in the ALIDA platform (Fig. 3). The dataset contains some categorical data, such as source and destination airports. It was used *string-indexer* to encode a string column of labels to a column of label indices service, assigning with '0' the most frequent label and so on in descending order, up to the number of labels. The data was clustered into '3' groups using a Spark K-Means application, in order to subsequently label them giving a meaning or relation to each group.

---

[7] OpenFlights/Airline Route Mapper Route Database - https://raw.githubusercontent.com/jpatokal/openflights/master/data/routes.dat.

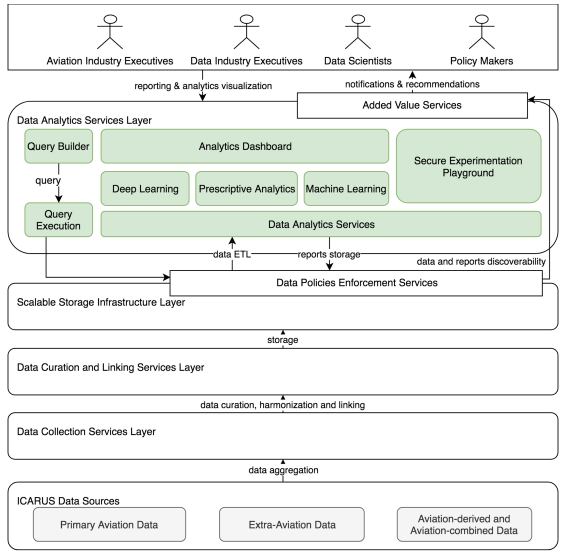

Figure 2. ICARUS Architecture.

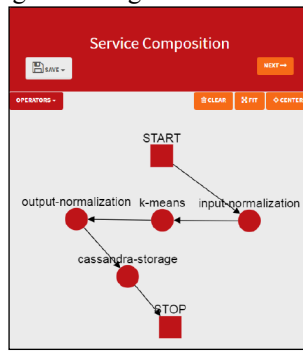

Figure 3. BDA applications workflow.

The Smart-Viz has been used to display the results in a scatter plot chart, shown in Fig. 4. This chart *sourceAirport* and *destinationAirport* fields colored by cluster. Another BDA application was run to have a summary of the dataset divided by clusters. The cluster '1' (green) is the one that includes the highest confluent of records, presenting a higher frequency of both source and destination airport. In the cluster '2' (red) there are, on average, records with less frequent source airports. Similarly, cluster '3' (blue) presents records with less frequent destination airports. This graphical analysis is supported by the summary obtained: about 80% of the records are associated to cluster '1'; moreover, routes with only direct flights are associated to cluster '2'. From the viewer it is possible to perform other queries to see how, for example, a specific airline is distributed, and plot the results in different charts available. This use case scenario has therefore shown how the ALIDA platform gives complete support for data analyst to perform BDA applications on data, and to extract the information from them.

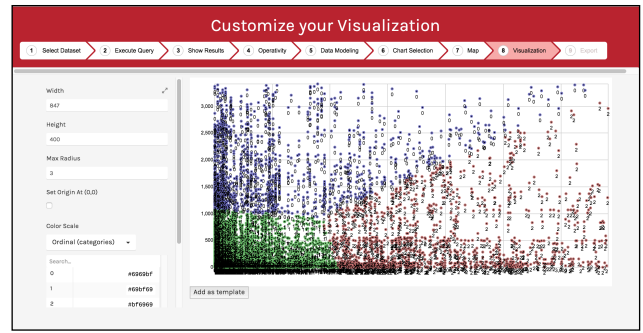

Figure 4. User scenario visualizazion.

## IV. CONCLUSIONS

ALIDA was presented as an innovative tool for the composition, deployment and execution of Big Data Analytics that makes use of the most cutting edge open source technologies. The ALIDA platform can be extended to support workflow management systems other than SCDF to handle different types of applications. Another future development could be to support the less skilled user also in the registration of a new service, through a user-friendly graphical user interface. This will allow to import, for example, a script to make it automatically compliant to the target platform. It is also our intention to extend the Configuration Optimizer to be able to manage an extensive set of metrics for several frameworks so to optimally deploy all BDA applications registered on the ALIDA service catalogue.

## V. ACKNOWLEDGMENTS

This work was partially supported by the EU's Horizon 2020 Innovation Programme through the ICARUS project (H2020-ICT-2016-2017/H2020-ICT-2017-1).

REFERENCES

[1] I. Hashem, I. Yaqoob, N. Anuar, S. Mokhtar, A. Gani, and S. Khan. The rise of big data on cloud computing: Review and open research issues. Information Systems, 47:98–115, 2015.

[2] Z. Zheng, J. Zhu, and M. Lyu. Service-generated big data and big data-as-a-service: an overview. In Proc. of BigData Congress 2013, Santa Clara, USA, June 2013.

[3] C.A. Ardagna, V. Bellandi, M. Bezzi, P. Ceravolo, E. Damiani, C. Herbert, "Model-based Big Data Analytics-as-a-Service: Take Big Data to the Next Level", in IEEE Transactions on Services Computing (TSC), 2018.

[4] C.A. Ardagna, P. Ceravolo, and E. Damiani. Big data analytics as-a service: Issues and challenges. In Proc. of PSBD 2016, Washington, VA, USA, December 2016.

[5] Guolu Wang, Jungang Xu1, Ben He, "A Novel Method for Tuning Configuration Parameters of Spark Based on Machine Learning", Sydney, NSW, Australia, High Performance Computing and Communications; IEEE 14th International Conference on Smart City; IEEE 2nd International Conference on Data Science and Systems (HPCC/SmartCity/DSS), 2016 IEEE 18th International Conference on, 2016.

[6] T. Ye and S. Kalyanaraman, "A recursive random search algorithm for large-scale network parameter configuration", Proc. of the International Conference on Measurements and Modeling of Computer Systems (SIGMETRICS 2003), San Diego, USA, June 9-14, 2013: 196-205.